\documentclass[pra,twocolumn,tightenlines,showpacs,nofootinbib]{revtex4}
\usepackage{bm,dcolumn,amsmath,graphicx,amsfonts,amssymb}
\usepackage{subfigure} 
\begin{document}
\title{Double core polarization contribution to atomic PNC and EDM calculations} 
\author{B. M. Roberts}
\author{V. A. Dzuba}
\author{V. V. Flambaum}
\affiliation{School of Physics, University of New South Wales, Sydney,
  New South Wales 2052, Australia} 
\date{ \today }


\begin{abstract}

We present a detailed study of the effect of the double core
polarization (the polarization of the core electrons due to the
simultaneous action of the electric dipole and parity-violating weak
fields) for amplitudes of the $ss$ and $sd$ parity non-conserving
transitions in Rb, Cs, Ba$^{+}$, La$^{2+}$, Tl, Fr, Ra$^{+}$,
Ac$^{2+}$ and Th$^{3+}$ as well as electron EDM enhancement factors
for the ground states of the above neutral atoms and Au. 
This effect is quite large and has the potential to resolve some
disagreement between calculations in the literature. It also has
significant consequences for the use of experimental data in
the accuracy analysis. 

\end{abstract}
\maketitle

\section{Introduction}

Measurements of parity nonconservation (PNC), and atomic electric
dipole moments (EDMs) provide important tests of the electroweak
theory (see, e.g.~the reviews~\cite{review2004,review2012}).  
The PNC amplitude of the $6s$-$7s$ transition in cesium is the most
precise low-energy test of the Standard Model to date.  
This precision is a result of the highly accurate measurements~\cite{meas} 
as well as the almost equally accurate atomic calculations~\cite{OurCsPNC2012,PorsevCs}, 
which are needed for their interpretation
(see also~\cite{CsPNC2002,CPM,rpaStrucRad}).  

For calculations of PNC in Cs there is very good agreement between
calculations, and the high accuracy is widely accepted. 
For other systems, however, there is disagreement between various
calculations -- sometimes by as much as 5\%.  
Due to the potential significance of these calculations for probing
physics beyond the Standard Model, it is very important that this
disagreement be resolved. 

In addition to the well known experiments for Cs, PNC measurements are
under consideration for the Ba$^+$ ion~\cite{BaII} and are in progress
for the Ra$^+$ ion~\cite{KVI}.  
The FrPNC collaboration has begun the construction of a laser cooling
and trapping apparatus with the purpose of measuring atomic parity
nonconservation in microwave and optical transitions of
francium~\cite{FrPNC}.  There are also experiments under way at the
Cyclotron and Radioisotope Centre (CYRIC) at Tohoku University to use
Fr in an electron EDM measurement~\cite{Kawamura2013}. For more
current and prospective EDM experiments see, e.g. \cite{review2004,review2012}.   

Reliable interpretation of all these measurements requires accurate
atomic calculations. In this paper we consider a particular aspect of
atomic calculations which has received little attention in previous
publications. This is the effect of the {double core polarization},
which arises due to the simultaneous action of the electric dipole (E1)
and parity-violating weak fields. The polarization of the atomic core
by the electric field of the laser is affected by the presence of the weak
interaction and vice versa. This leads to an additional contribution to
the PNC amplitude or atomic EDM, which varies significantly
between different atoms and transitions. It is 0.26\% for the $6s$-$7s$ PNC transition in Cs but is significantly larger for the $sd$
PNC transitions -- reaching 6\% for the $6s$-$5d_{3/2}$ PNC transition
in Ba$^+$. A special case is the thallium atom. If thallium is treated as
a mono-valence system then the double core
polarization contribution is about 40\% for the $6p_{1/2}$-$6p_{3/2}$
PNC amplitude and about 60\% for the EDM induced in the 6$p_{1/2}$ ground state.

The importance of the double core polarization contribution is known
and is included in many PNC and EDM calculations (see,
e.g. \cite{dzuba1987,LiuKelly92}). However, it was never studied in
detail and its importance was never properly emphasized. 
Perhaps for this reason it may be that some calculations based on the sum-over-states approach have missed this contribution. 

In this work we study the effect of the double core polarization for
the amplitudes of the $ss$ and $sd$ parity non-conserving
transitions in Rb, Cs, Ba$^{+}$, La$^{2+}$, Tl, Fr, Ra$^{+}$,
Ac$^{2+}$ and Th$^{3+}$ as well as electron EDM enhancement factors
for the ground states of the above neutral atoms and Au. We show that
the effect is large and in some cases can explain the discrepancy 
between different calculations. We also show that this contribution
affects the analysis of the accuracy of the calculations based on the
use of the experimental data.

\section{Calculations}

\subsection{PNC and EDM amplitudes}

The PNC  amplitude, $E_{PNC}$, of a transition  between states of the
same parity can be expressed via the sum over all possible
intermediate opposite parity states $n$,  
\begin{equation}%
\small
E_{PNC} = \sum_n 
	 \Big[ \frac{\langle
 b|\hat{d}_{E1}|{n}\rangle\langle{n}|\hat{h}_{PNC}|{a}\rangle}{\varepsilon_{a}
 - \varepsilon_{n}} + \frac{\langle b|\hat{h}_{PNC}|n\rangle\langle
          n|\hat{d}_{E1}|{a}\rangle}{\varepsilon_{b}-\varepsilon_{n}}\Big], 
	\label{eq:pncsum}
\end{equation}
where $a$, $b$, and $n$ are many-electron wave functions of the atom
with corresponding energies $\varepsilon$, $\hat{d}_{E1}$ is the
electric dipole transition operator and  $\hat{h}_{PNC}$ 
is the operator of the weak interaction.

Likewise, the contribution to an atomic EDM induced in the atomic
state $a$ by  a mixing of opposite parity states $n$ has the form 
\begin{equation}%
 {d}_\text{atom} =2 \sum_n 
\frac{\langle
 a|\hat{d}_{E1}|{n}\rangle\langle{n}|\hat{h}_{PT}|{a}\rangle}{\varepsilon_{a}
  -\varepsilon_{n}}   
\label{eq:edmsum},
\end{equation}
where $\hat{h}_{PT}$ is the $P$- and $T$-odd operator that depends on
the electron EDM and mixes states of opposite parity.

 The amplitudes can then be evaluated via a direct summation of
 products of matrix elements and energy denominators over the states $n$.  
We refer to this method as the {direct-summation} (DS) method.
We, however, bypass this technique in favour of a more numerically
stable approach based on solving differential equations, the so-called
{solving-equations} (SE) method, or the mixed-states method. 
This approach, which is outlined in the next section, has many
important advantages, not least of which is that it allows the easy
inclusion of the important {double core polarization} (DCP)
contribution. 

\subsection{Atomic structure calculations and core polarization}

The above `exact' expressions~(\ref{eq:pncsum}) and (\ref{eq:edmsum})
can be reduced to approximate formulas containing instead
single-electron energies and matrix elements. Then many-body effects
are included by modifying the single-electron orbitals and the
external field operators.  

We begin with the relativistic Hartree-Fock approximation and proceed
to include the dominating electron correlation effects using the
correlation potential method~\cite{CPM}. 
The correlation potential is used to construct the so-called {Brueckner orbitals} (BOs) for the valence electron, which are found
by solving the Hartree-Fock-like equations with the extra operator
$\hat \Sigma$:   
\begin{equation}
 (\hat H_0 +\hat \Sigma - E_n)\psi_n^{(\rm BO)}=0,
\label{eq:BO}
\end{equation}
where $\hat H_0$ is the relativistic Hartree-Fock Hamiltonian and the
index $n$ denotes valence states. The BO $\psi_n^{(\rm BO)}$ and
energy $E_n$ include correlations.   

Interactions with the external fields are included via the
time-dependent Hartree-Fock  (TDHF) approximation (see, e.g.~\cite{CPM,RPA}).  
The external fields in question are the electric dipole (E1)
interaction with the electric field of the photon, $d_{E1}$, and
either the  nuclear-spin--independent weak interaction $h_{PNC}$, or
the P- and T-odd weak interaction $h_{PT}$ in the case of atomic EDMs.  
It is with this method that we also include the important
core-polarization effects, which arise from the action of the external
fields on the Hartree-Fock $V^{N-1}$ core potential.  

Within the framework of the TDHF method, the single-electron
wavefunction in external weak and $E1$ fields is expressed  
\begin{equation}
\psi = \psi_0 + \delta\psi +X e^{-i\omega t}+Y e^{i\omega t} + \delta
Xe^{-i\omega t} + \delta Ye^{i\omega t},
\end{equation}
where $\psi_0$ is the unperturbed state, $\delta\psi$ is the
correction due to the weak interaction acting alone,  $X$ and $Y$ are
corrections due to the photon field acting alone, $\delta X$ and
$\delta Y$ are corrections due to both fields acting simultaneously,
and $\omega$ is the frequency of the PNC transition. 
Since the EDM amplitude is a diagonal matrix element with no
transition, $\omega=0$ in the EDM case. 
This method is equivalent to the well-known {random phase approximation} (RPA).



%
%


The corrections $\delta V$ to the core potential are found by solving
the following system of RPA equations self-consistently for the core
states.  

The equations for the E1 core polarization
\begin{equation}
\begin{aligned}
	\label{eq:rpaE1}
(\hat H_0 - E_c-\omega)X_c
	&= - (\hat d_{E1} + \delta \hat 	V_{E1})\psi_{0c}, 
	 \\ 
(\hat H_0 - E_c+\omega)Y_c
	&= - (\hat d_{E1}^{\dagger} + \delta \hat V_{E1}^{\dagger})\psi_{0c}, 
\end{aligned}
\end{equation}
and for the weak core polarization
\begin{equation}
\begin{aligned}
	\label{eq:rpaW}
(\hat H_0 - E_c)\delta\psi_c
	&= - (\hat f + \delta \hat V_f)\psi_{0c},
\end{aligned}
\end{equation}
are independent and can be solved separately.
Here, the index $c$ denotes core states, $\hat f$ is the operator of
the external weak field, and $\delta \hat V_f$ and $\delta \hat
V_{E1}$ are corrections to the core potential arising from the weak
and $E1$ interactions respectively. Again, $\omega$ is the energy of the PNC transition, and is zero in the case of EDMs.

There is also the set of equations corresponding to the double core
polarization: 
\begin{eqnarray}
	\label{eq:rpaDCP}
(\hat H_0 - E_c-\omega)\delta X_c  & = -\delta\hat V_{E1}\delta\psi_c - \delta\hat V_f X_c  \nonumber \\
	& -  \delta\hat V_{fE1}\psi_{0c} + \delta E_c\psi_{0c} ,  \\ 
(\hat H_0 - E_c+\omega)\delta Y_c   &=   
       -\delta\hat V_{E1}^{\dagger}\delta\psi_c -
	\delta\hat V_fY_c  \nonumber \\
&- \delta\hat V_{fE1}^{\dagger}\psi_{0c} +   \delta E_c\psi_{0c}. \nonumber
\end{eqnarray}
Here, $\delta \hat V_{fE1}$ is the correction to the core potential
arising from the simultaneous perturbation of the weak field and the
electric field of the laser light, and $\delta E_c$ is the
corresponding correction to the core energy. 
The correction to the core energy, $\delta E_c$, 
is zero in the case of PNC, but non-zero for EDMs (see equation~(\ref{eq:seedm})).

The equations (\ref{eq:rpaDCP})  depend on the solutions to equations
(\ref{eq:rpaE1}) and (\ref{eq:rpaW}), and must therefore be iterated
after (\ref{eq:rpaE1}) and (\ref{eq:rpaW}) are solved.  
In the solving-equations method, the PNC amplitude between valence
states $a$ and $b$ is then given by 
\begin{equation}
\begin{aligned}
	 E_{\rm PNC} 	
&= \langle \psi_b|\hat d_{E1}
		 +  \delta\hat V_{E1}|\delta\psi_a\rangle \\
		&+ \langle \psi_b|\hat h_{PNC} + \delta\hat V_{W}|X_a\rangle  
		+ \langle \psi_b|\delta\hat V_{fE1}|\psi_a\rangle \label{eq:sepnc}  \\
	&= \langle \psi_b|\hat d_{E1} + \delta\hat V_{E1}|\delta\psi_a\rangle \\
		&+  \langle \delta\psi_b|\hat d_{E1} + \delta\hat V_{E1}|\psi_a\rangle  
		+ \langle \psi_b|\delta\hat V_{fE1}|\psi_a\rangle,
\end{aligned}
\end{equation}
and the corresponding atomic EDM is given by
\begin{equation}
	d_{\rm atom} = 2 \langle \psi_a|\hat d_{E1} + 
	 \delta\hat V_{E1}|\delta\psi_a\rangle +
	\langle \psi_a|\delta\hat V_{fE1}|\psi_a\rangle.
	 \label{eq:seedm} 
\end{equation}
By using BOs for the valence states $\psi_a$ and $\psi_b$ in
(\ref{eq:sepnc}) and (\ref{eq:seedm}) we can include correlations in
the calculation of the PNC and EDM amplitudes. The corrections $\delta
\psi_a$ and $\delta \psi_b$ to the BOs $\psi_a$ and $\psi_b$ are also
found with the use of the correlation potential $\hat \Sigma$: 
\begin{equation}
(\hat H_0 -E_a + \hat \Sigma)\delta\psi_a = -(\hat h_f + \delta\hat V_f)\psi_{0a}. \label{eq:dpsiv} 
\end{equation}

The last term in equations (\ref{eq:sepnc}) and (\ref{eq:seedm})
represents the double core polarization contribution (DCP), which is due to the simultaneous action of the two external fields.  
This term gives an important correction that is often not included in
sum-over-states calculations. 

It is possible to include a term for the DCP perturbatively 
directly after solving
equations~(\ref{eq:rpaE1}) and (\ref{eq:rpaW}) and without iterating
the equations~(\ref{eq:rpaDCP}).  
This contribution corresponds to the lowest order DCP term, which we
refer to as $\delta V_{fE1}^{\rm pert.}$. 
There is, however, another contribution that comes from further
iterations of the pair of equations~(\ref{eq:rpaDCP}).  
This effect, which we refer to as the relaxation effect $\delta
V_{fE1}^{\rm relax}$, has a significant impact on the value of the
double core polarization. 
The relative size of this relaxation effect means it is not enough to
simply include the term perturbatively, and the total DCP term must be
taken as $\delta V_{fE1}=\delta V_{fE1}^{\rm pert.}+\delta
V_{fE1}^{\rm relax}$.

In these calculations we didn't include corrections such as structure radiation 
(the correction to the correlation potential $\Sigma$ due to the E1 field, $\delta\Sigma_{E1}$, the weak correlation potential, $\delta\Sigma_{W}$, and the combined weak and E1 fields, $\delta\Sigma_{WE1}$), or other higher order corrections such as ladder diagrams, and renormalization of states.
These corrections are typically small (with perhaps the exception of thallium when treated as a single valence system) though they should be taken into account for accurate calculations.

\section{Results and discussion}

\subsection{PNC amplitudes}

We have performed calculations of the double core polarization
correction to many PNC amplitudes, the results of which are presented
in Table~\ref{tab:pnc} along with several existing PNC calculations
for comparison. 
We present  the contributions of the double core polarization that
stem from including the term perturbatively, $\delta V_{fE1}^{pert.}$,
and the subsequent relaxation effect, arising from further iterations
of (\ref{eq:rpaDCP}), $\delta V_{fE1}^{relax}$, separately.

\begin{table*}
  \caption{Double core polarization contribution to parity
    nonconservation amplitudes for transitions in several atoms and
    ions. 
We present several of the most complete calculations, and what their
value would be if the DCP term was omitted ($E_{PNC}^{-\delta
  V_{fE1}}$). 
Shown separately are  the lowest order perturbative DCP term, $\delta
V_{fE1}^{\rm pert.}$, and the {relaxation} contribution that comes
from iterations of the equations (\ref{eq:rpaDCP}), $\delta
V_{fE1}^{\rm relax}$. 
Also shown are several available calculations and the methods they
used for comparison.  
SE refers to the solving-equations (or mixed-states) method, which
typically includes the DCP term, and DS is the direct-summation
method, which typically doesn't. 
Amplitudes are presented in units of $iea_B (-Q_W/N)\times10^{-11}$.}
\begin{ruledtabular}
  \begin{tabular}{lldldddddlll}
 \multicolumn{4}{c}{$E_{PNC}$}   &   \multicolumn{4}{c}{DCP contribution - This work}   &   \multicolumn{4}{c}{Other values} \\
\cline{1-4}\cline{5-8}\cline{9-12}\\
\multicolumn{2}{c}{Transition}   &   \multicolumn{2}{c}{Most complete}   &   \multicolumn{1}{c}{$\delta V_{fE1}^{pert.}$}   &   \multicolumn{1}{c}{$\delta V_{fE1}^{relax}$}   &   \multicolumn{1}{c}{\% $\delta V_{fE1}^{total}$}   &   \multicolumn{1}{c}{$E_{PNC}^{-\delta V_{fE1}}$}   &   \multicolumn{2}{c}{$E_{PNC}$}   &   \multicolumn{2}{c}{Method}\\   
\hline\\
$^{85}$Rb   &   $5s$-$6s$   &   0.1390(7) &   \cite{OurRbPNC}   &   -0.0004   &   0.0001   &   -0.24\%   &   0.1393   &   0.139(2)    &   \cite{dzuba1987}   &   SE   &   MBPT\\
   &   $5s$-$4d_{3/2}$   &   -0.450   &     
    &   0.0065   &   0.0021   &   -2.0\%   &   -0.459   &   \multicolumn{1}{c}{---}   &   \multicolumn{1}{c}{}   &   \multicolumn{1}{c}{}   &   \multicolumn{1}{c}{}\\
$^{133}$Cs   &   $6s$-$7s$   &   0.9041(45)   &   \cite{CsPNC2002}   &   -0.0034   &   0.0010   &   -0.26\%   &   0.907   &   0.8977(40)   &   \cite{OurCsPNC2012}   &   DS   &   MBPT   \\ 
   &      &      &      &      &      &      &      &   0.8906(24)   &   \cite{PorsevCs}   &   DS   &   CC\\
   &   $6s$-$5d_{3/2}$   &   -3.70(4)   &   \cite{Fr-like}   &   0.070   &   0.030   &   -2.6\%   &   -3.80   &   -3.76(7)   &   \cite{Dzuba2001}   &   DS   &   MBPT\\
   &      &      &      &      &      &      &      &   -3.62(7)   &   \cite{Dzuba2001}   &   SE   &   MBPT\\
$^{137}$Ba$^+$   &   $6s$-$7s$   &   0.658(7)   &   \cite{Fr-like}   &   -0.007   &   0.001   &   -0.84\%   &   0.664   &   \multicolumn{1}{c}{---}   &   \multicolumn{1}{c}{}   &   \multicolumn{1}{c}{}   &   \multicolumn{1}{c}{}\\
   &   $6s$-$5d_{3/2}$   &   -2.20(2)   &   \cite{Fr-like}   &   0.073   &   0.067   &   -6.0\%   &   -2.34   &   -2.34(9)   &   \cite{Dzuba2001}   &   DS   &   MBPT\\
   &      &      &      &      &      &      &      &   -2.17(9)    &   \cite{Dzuba2001}   &   SE   &   MBPT\\
   &      &      &      &      &      &      &      &   -2.46(2)    &   \cite{SahooBa+}   &   DS   &   CC\\
$^{139}$La$^{2+}$   &   $6s$-$5d_{3/2}$   &   -2.14(2)   &   \cite{Fr-like}   &   0.051   &   0.085   &   -6.0\%   &   -2.28   &   \multicolumn{1}{c}{---}   &   \multicolumn{1}{c}{}   &   \multicolumn{1}{c}{}   &   \multicolumn{1}{c}{}\\
$^{223}$Fr   &   $7s$-$8s$   &   15.49(16)   &   \cite{ShabFr2005}   &   -0.05   &   0.05   &   -0.06\%   &   15.5   &   15.41   &   \cite{Safronova2000}   &   DS   &   CC\\
   &      &      &      &      &      &      &      &   15.9(2)   &   \cite{Dzuba1995}   &   SE   &   MBPT\\
   &   $7s$-$6d_{3/2}$   &   -58.0(6)   &   \cite{Fr-like}   &   1.12   &   0.40   &   -2.6\%   &   -59.5   &   -59.5(24)    &   \cite{Dzuba2001}   &   DS   &   MBPT\\
   &      &      &      &      &      &      &      &   -57.1(23)    &   \cite{Dzuba2001}   &   SE   &   MBPT\\
$^{226}$Ra$^+$   &   $7s$-$8s$   &   10.9(1)   &   \cite{Fr-like}   &   -0.10   &   0.07   &   -0.28\%   &   10.9   &   \multicolumn{1}{c}{---}   &   \multicolumn{1}{c}{}   &   \multicolumn{1}{c}{}   &   \multicolumn{1}{c}{}\\
   &   $7s$-$6d_{3/2}$   &   -44.3(4)   &   \cite{Fr-like}   &   1.29   &   0.92   &   -4.8\%   &   -46.5   &   -45.89  &   \cite{SafRa+2009}   &   DS   &   CC\\
   &      &      &      &      &      &      &      &   -46.4(14)   &   \cite{SahooR+}   &   DS   &   CC\\
   &      &      &      &      &      &      &      &   -45.9(19)   &   \cite{Dzuba2001}   &   DS   &   MBPT\\
   &      &      &      &      &      &      &      &   -43.9(18)\tablenotemark[1]   &   \cite{Dzuba2001}   &   SE   &   MBPT\\
$^{227}$Ac$^{2+}$   &   $7s$-$6d_{3/2}$   &   -42.8(4)   &   \cite{Fr-like}   &   1.01   &   1.21   &   -4.9\%   &   -45.0   &   \multicolumn{1}{c}{---}   &   \multicolumn{1}{c}{}   &   \multicolumn{1}{c}{}   &   \multicolumn{1}{c}{}\\
$^{232}$Th$^{3+}$   &   $7s$-$6d_{3/2}$   &   -43.6(4)   &   \cite{Fr-like}   &   0.75   &   1.44   &   -4.8\%   &   -45.8   &   \multicolumn{1}{c}{---}   &   \multicolumn{1}{c}{}   &   \multicolumn{1}{c}{}   &   \multicolumn{1}{c}{}\\
  \end{tabular}%
\end{ruledtabular}%
\tablenotetext[1]{Rescaled from $^{223}$Ra$^+$}%
  \label{tab:pnc}%
\end{table*}%


Our results show that the double core polarization term is quite
large, especially for the $sd$ PNC transitions, and also that the
relaxation effect is not small and must be included along with the
perturbative lower order term.  
We also demonstrate that in these cases the majority of the
discrepancy between the solving-equations (SE) and direct
sum-over-states (DS) calculations can be explained by the possible omission of the DCP term. 

In Ref.~\cite{Dzuba2001},  calculations of $sd$ PNC transitions  were performed for Cs,
Ba$^+$, Fr and Ra$^+$ using both the solving-equations and the
direct-summation methods. 
 As discussed, the double core polarization contribution was included
 in \cite{Dzuba2001} in the SE calculations only. 
In that work there was about a $4\%$ discrepancy between the DS and SE
calculations for Cs and Fr, 8\% for Ba$^+$ and 7\% for Ra$^+$.  
Here, we calculate the contribution of the double core polarization
for these $s$-$d$ transitions to be approximately $3\%$ for Cs and Fr, 6\% for
Ba$^+$ and 5\% Ra$^+$ -- consistently making up for most of the
disagreement. 
The rest of the difference likely comes from the numerical accuracy of
the different methods and minor differences in correlation
calculations.  
If the double core polarization contribution is removed from the SE
calculations then our SE and DS calculations
match perfectly for Ba$^+$ and Fr, and are within 1\% for Cs and
Ra$^+$~\cite{Dzuba2001}.  


The $sd$ transition in Ra$^+$ is a particularly useful case to
study as there are a number of values available for comparison. Total
DCP contribution is about $-5$\% (see Table \ref{tab:pnc}) which is very
close to the difference between the most complete calculations of
Ref.~\cite{Fr-like} and all calculations using the DS approach where
this contribution may be missing.
The range of values that do not include the double core polarization
term, including the DS values from Ref.~\cite{Dzuba2001}, lie within 1\% of each other.  
They also lie within 1\% of the value obtained by removing the DCP
contribution from the result of Ref.~\cite{Fr-like}.

Another value, calculated by Wansbeek {\em et al.}~\cite{SahooR+}
using a relativistic coupled-cluster (CC) approach, also agrees
with these values, lying within 0.3\% of the value calculated in this
work without double core polarization and 0.2\% of the Pal {\em et
  al.}~\cite{SafRa+2009} DS value.  
We are not sure if the DCP contribution was included in the works~\cite{SafRa+2009,SahooR+}.

The difference between the $6s$-$7s$ PNC transitions in Cs for the
solving-equations value  0.9041(45) of Ref.~\cite{CsPNC2002} and the
sum-over-states value 0.8906(24) of Ref.~\cite{PorsevCs} is larger
than the DCP term -- it is mainly due to missed contributions to the
core and tail parts of the summation in (\ref{eq:pncsum})
(see~\cite{OurCsPNC2012} for full detail). 
It is worth noting however, that the double core polarization
contribution of $0.26\%$ is of the same size as the uncertainty quoted in~\cite{PorsevCs} of
0.27\% -- meaning that this uncertainty can only be claimed if the DCP
contribution is included. 
As we shall discuss in the next section, the double core polarization
contribution has particular impact on the accuracy analysis.


We have performed detailed PNC calculations for these Fr- and Cs-like
ions in our recent paper, Ref.~\cite{Fr-like}.  
A more complete analysis of the accuracy of these calculations,
including calculations of energy levels, lifetimes and matrix elements
is given in that work.

\subsection{Atomic EDM}

As well as parity nonconservation, calculations for several atomic
electric dipole moments (EDMs) induced by the dipole moment of the
electron ($d_e$) have been performed. 
These calculations, along with several existing calculations for
comparison, are presented in Table~\ref{tab:edm}.  

Our previous calculations of the EDM for Cs~\cite{Dzuba2009a}, as well
as Fr and Au~\cite{dzuba1999}, do not include the double core
polarization term.  
These values, along with one for Rb calculated in this work, are
presented in the $d_{atom}^{(0)}$ column of Table~\ref{tab:edm}.  
They  are then corrected by adding the DCP term with the corrected
results given in the column $d_{atom}^{\rm new}$. 
We find here also that the double core polarization term is quite a
large contribution, and that by including this term we can improve the
agreement between our previous values and several other calculations. 

It is interesting to note that if we include only the perturbative DCP
term into the EDM calcualtions for Cs and Fr and don't include the
relaxation term, we reproduce the values from References~\cite{SahooRbCs} and~\cite{sahooFr2009} almost exactly (see Table~\ref{tab:edm}).

  \begin{table*}
    \caption{
Double core polarization contribution to Atomic EDM calculations for several atoms including both the perturbative and relaxation parts. 
The values $d_{atom}^{(0)}$ do not include DCP, and the values $d_{atom}^{\rm new}$ do.
Values in units of  $d_e$.} 
\begin{ruledtabular}  
  \begin{tabular}{llldddldl}%
 \multicolumn{2}{c}{}    &        
 \multicolumn{1}{c}{}    &   
 \multicolumn{4}{c}{DCP – This work}    &       
\multicolumn{2}{c}{Other values} \\   
\cline{4-7}\cline{8-9}\\
 \multicolumn{2}{c}{State}    &     
\multicolumn{1}{c}{$d_{atom}^{(0)}$}    &   
\multicolumn{1}{c}{$\delta V_{fE1}^{\rm pert.}$}   & 
\multicolumn{1}{c}{$\delta V_{fE1}^{\rm relax.}$}   & 
\multicolumn{1}{c}{$\%\,\delta V_{fE1}^{\rm both}$}     &    
\multicolumn{1}{c}{$d_{atom}^{\rm new}$}    &    
\multicolumn{1}{c}{$d_{atom}$}    &   
\multicolumn{1}{c}{Ref.} \\
\hline  
Rb      &      $5s$      &   26.8\tablenotemark[1]   &   -0.59   &   -0.86   &      -5.4\%      &   25.4   &   25.74(26)   &      \cite{SahooRbCs}   \\
      &            &      &            &            &            &      &   25.7   &      \cite{PRA1994}   \\
      &            &      &            &            &            &      &   24.6   &      \cite{JohnsonEDM86}   \\
Cs      &      $6s$      &   124(4)\tablenotemark[2]   &   -3.0   &   -2.5   &      -4.4\%      &   119(4)   &   120.5(12)   &      \cite{SahooRbCs}   \\
      &            &      &            &            &            &      &   114.9   &      \cite{JohnsonEDM86}   \\
Au     &      $6s$      &   260(39)\tablenotemark[3]   &   -6.7   &   -3.4   &      -3.9\%      &   250(39)   &   249.9   &      \cite{JohnsonEDM86}   \\
Fr      &      $7s$      &   910(46)\tablenotemark[3]   &   -24.3   &   -12.1   &      -4.0\%      &   874(46)   &   894.93   &      \cite{sahooFr2009}   \\
  \end{tabular}%
\end{ruledtabular}
\tablenotetext[1]{This work.}
\tablenotetext[2]{Reference~\cite{Dzuba2009a}.}
\tablenotetext[3]{Reference~\cite{dzuba1999}.}
    \label{tab:edm}%
  \end{table*}%

The thallium atom represents an interesting case for both PNC and EDM
calculations. If we treat Tl as a mono-valence system, then the DCP
contribution to the PNC amplitude is huge. It contributes 36\% to the PNC amplitude
of the $6p_{1/2}$ - $6p_{3/2}$ transition and about 60\% to the EDM of
the ground state. The DCP contribution is strongly dominated by the
$6s$ electrons. This reflects the well-known fact that the correlations
between three outermost electrons in thallium are strong and should be
treated accurately. In our view, the best approach is to treat
thallium as a triple-valence-electron system and to use the
configuration interaction (CI) technique combined with many-body perturbation theory (MBPT) for including valence electrons core interactions. 
However, good results can be
obtained in other approaches too if correlations between $6s$ and $6p$
electrons, including the DCP contribution, are treated accurately. In
our early calculations of the PNC in thallium~\cite{dzuba1987}
it was treated as a mono-valence system and the DCP contribution was
included. Recent
calculations of the EDM enhancement
factor~\cite{Dzuba2009a,kozlov2012} used the CI approach,
the calculations of the Tl EDM based on the coupled-cluster approach
~\cite{LiuKelly92,SahooTlEDM} seems to include the DCP contribution too by
introducing the perturbed excitation operators $T_1$ and $T_2$ (see
\cite{LiuKelly92} for detailes).

\section{Implications to accuracy analysis}\label{sec-accuracy}

Most of the accuracy analysis in the literature assumes that the PNC
and EDM amplitudes can be reduced to a sum of products of matrix
elements and energy denominators that are all independent.  
The E1 matrix elements and energies can then be compared with
experimental values in order to judge the accuracy of the
calculations. 
The accuracy of the weak matrix elements can similarly be judged by
calculating hyperfine structure constants, since both the weak
interaction and the hfs rely on the form of the wavefunctions on short
distances. 
The accuracy of this analysis, however, is limited by the value of the
double core polarization effect -- which is by no means negligible. 
The DCP contribution cannot easily be presented as a product of weak and
electric dipole matrix elements which are independent on each
other. If the analysis of accuracy ignores this contribution it does not present the whole picture.



\section{Conclusion}

We have calculated the contribution of the double core polarization
effect to the PNC and EDM amplitudes of several atomic systems.   
This is an important contribution that is of the same order or even
larger than the Breit~\cite{breit}, neutron-skin~\cite{neutronskin},
and QED~\cite{QED} corrections. 
This term  has the potential to restore the agreement between
differing calculations.

\paragraph*{Acknowledgments---}
This work was supported by the Australian Research Council.%


\end{document}